\documentclass{elsart}
\usepackage{graphicx}

\journal{Journal of Computational Physics}

\def\be{\begin{equation}}
\def\ee{\end{equation}}

\begin{document}
\begin{frontmatter}
\title{Numerical solution of perturbed Kepler problem using a split operator technique}
\author[Goutham]{G. S. Balaraman}
\author[Daniel]{and D. Vrinceanu}
\ead{vrinceanu@lanl.gov}
\address[Goutham]{School of Physics, Georgia Institute of Technology, Atlanta, GA 30332, USA}
\address[Daniel]{Theoretical Division, Los Alamos National Laboratory, Los Alamos, NM 87545, USA}

\begin{abstract}
An efficient geometric integrator is proposed for solving the perturbed Kepler motion. This
method is stable and accurate over long integration time, which makes it appropriate for
treating problems in astrophysics, like solar system simulations, and atomic and molecular
physics, like classical simulations of highly excited atoms in external fields. The key idea
is to decompose the hamiltonian in solvable parts and propagate the system according to each term.
Two case studies, the Kepler atom in an uniform field and in a monochromatic field, are presented and
the errors are analyzed. 
\end{abstract}
\date{\today}
\end{frontmatter}

\section{Introduction}
The motion of a body under a force which depends inverse quadratically with the distance is known as the
Kepler problem and explains both the planetary dynamics and the electronic structure of atoms. The problem
is completely solvable and is a textbook example of constructing action-angle variables for a nontrivial
system. The stable computation of trajectories of perturbed Kepler systems is important for astronomical
applications, such as solar system studies, and in the classical and semi-classical studies of atomic
systems, particularly atoms in highly excited (Rydberg) states. Once the initial conditions are assigned,
the resulting system of ordinary differential equations can be numerically solved using standard,
good-for-all, algorithms such as Runge-Kutta or Gear integrators. Although such methods have a good control
over the local error, their global error usually grows exponentially. This is why the traditional numerical
methods are unsuitable in treating physical problems which require long time integration.
The simulation of the solar system for more than $10^9$ years \cite{Laskar1994} is an example
where special integration methods are needed \cite{Wisdom1991}.
Due to the weak electromagnetic coupling (of the order of fine structure constant cubed $\approx 1/137^{3}$)
the simulation of the interaction of a Rydberg atom with radiation also requires following the electron
trajectory for a long time.

Geometric integration is a relatively new branch of the Computational Mathematics. It studies algorithms and
discretization methods that respect the underlying geometry and qualitative structure of the the problems it
aims to solve. The principle is that if more specific information about the problem is explicitly 
included into the solver, then the solution will be more accurate and stable than those produced by generic methods.
As a simple illustration, consider the harmonic oscillator problem
$
du/dt = v, \; dv/dt = -u
$.
This system has the property that (a) its solutions are all periodic and, (b) bounded and, (c) that 
$u^{2}+v^{2}$ is a conserved quantity of the evolution.
A forward Euler discretization of this system with time step $h$ gives
$
u_{n+1} = u_{n} + h v_{n}, \; v_{n+1}=v_{n}-h u_{n}
$.
It is easy to see that
$
u_{n+1}^{2} + v_{n+1}^{2} = (1+h^{2})(u_{n}^{2}+v_{n}^{2})
$.
While the correct solution is obtained in the limit of $h\rightarrow 0$, the numerical method, over any finite time
interval, has lost all three qualitative features listed above!

For a Hamiltonian system, a numerical method is {\it symplectic} if it produces an approximate solution and in the same
time it explicitly preserves the underlying symplectic structure.
Many symplectic methods also obey other qualities of the system such as
symmetry and time reversal. Although the Hamiltonian itself is not conserved in general 
(for autonomous systems), it was demonstrated that the error grows only linearly in time. Comprehensive
reviews of symplectic methods are presented in references \cite{Leimkuhler2005} and \cite{Hairer2002}. 

Suppose that the Hamiltonian is separable and each term in the separation defines an integrable problem,
then a symplectic solution is obtained by propagating the system with each term separately.
The symplectic quality is evident since the composition of two symplectic maps is also
symplectic and the solution is exact in the limit of vanishing time step.
The traditional way is to split the Hamiltonian as the
kinetic energy $T$ plus potential energy $V$.
Evolution under the kinetic energy part alone describes the free motion of the system.
The potential energy term usually does not depend on momentum coordinates and therefore
the position coordinates remain unchanged during propagation with it. The system gets a "kick", a
sudden change in momentum.
It was proved (see \cite{Leimkuhler2005}, for example) that the leap-frog, or St\"ormer-Verlet, methods, very popular
in Molecular Dynamics simulations, can be derived from a $T+V$ splitting and are therefore symplectic.
Essential for a more general way of splitting the Hamiltonian is that the parts are not only integrable 
but also efficiently computable.

This paper proposes a symplectic integration method for perturbed Kepler problems by splitting the Hamiltonian
into a pure, exactly solvable, Kepler part and a perturbation part.
Even when the perturbation is not small, the system can be propagated with accuracy using relatively large
time steps.
This method can be applied for all problems where such a decomposition
makes sense. For a collision problem, such as electron scattering on an atom, this splitting is not practical
when the projectile and target electrons come close to each other and interact strongly. In such a case the time
step has to be chosen very small so that trajectories are straight lines (during one time step) and this
method is no more efficient than the simpler $T+V$ splitting.
Although the algorithm is still correct, the advantage of being able to
take large time steps is lost.
 
Two case studies are presented to demonstrate the efficiency of the proposed method.
Interaction of a Rydberg atom with an uniform external field is studied in both cases.
One case treats the interaction with a constant field (Stark effect)
and the other with that of a monochromatic oscillatory field (laser interaction).

\section{Symplectic propagator}
The motion of a planet or of a Rydberg electron can be described by the classical equations of motion
derived from the Hamiltonian

\[
H = H_{0} + V 
\]
where $H_{0} = p^{2}/2 - 1/r$ (in an appropriate set of units) is the Kepler Hamiltonian and V is 
the perturbation potential. The evolution in time of any function $z$ defined over the phase
space is given by the general equation
\[
\frac{dz}{dt} = \{ z, H\} \equiv D_{H}(z)
\]
where $\{\;,\;\}$ are the Poisson brackets and $D_{H}$ is an operator defined by the Hamiltonian $H$. 
Providing the initial condition $z_{0}$ at $t_{0}$, the function $z$ is obtained at time $t$ as a solution
of the above equation, given as $z(t) = \Phi_{\Delta t}(z_{0})$. Formally, the mapping $\Phi$ can written as
$\Phi_{\Delta t} = e^{\Delta tD_{H}}$, in a form reminiscent of the quantum mechanical evolution operator.
Of course, the exponential of operator $D_{H}$ has an exact meaning only if the solution
of the Hamiltonian $H$ is explicitly known. Apparently no real advantage is gained.
However, solutions for $H_{0}$ and $V$ are
separately known, and, for any $\Delta t$, the mapping associated with these solutions are
$e^{\Delta t D_{H_{0}}}$ and $e^{\Delta t D_{V}}$. If the phase space operators $D_{H_{0}}$ and $D_{V}$ 
commute (or their Poisson bracket $\{H_0,V\}=0$) then the product of exponentials is equal to the
exponential of their sum, and the problem has an exact solution: 
$\Phi_{\Delta t} = e^{\Delta t D_{H_{0}}} e^{\Delta t D_{V}}$.
This is not the case in general, and
the following useful expansion can be derived from the Baker-Campbell-Hausdorff formula
\begin{eqnarray}
\log (e^{\lambda A /2}e^{\lambda B}e^{\lambda A/2}) &=& 
\lambda (A + B) - \frac{\lambda^{3}}{24}[AAB] - \frac{\lambda^{3}}{12}[BAB] \label{BCH}\\
&& +\frac{7 \lambda^{5}}{5760}[AAAAB] + \frac{\lambda^{5}}{720}[BBBAB] \nonumber\\
&& +\frac{\lambda^{5}}{480}[AABAB] + \frac{\lambda^{5}}{3600}[BAAAB] \nonumber\\
&& -\frac{\lambda^{5}}{360}[ABBAB] + \frac{\lambda^{5}}{120}[BABAB]  + {\cal O}(\lambda ^{7})\nonumber
\end{eqnarray}
where the bracket notation refers to commutators: $[XY] = [X,Y] = AB - BA$, $[XYZ] = [X,[Y,Z]]$, 
and so on. Operators $A$ and $B$ can be either $D_{H_{0}}$ or $D_{V}$. The Poisson bracket also has
the property of being a Lie brackets which means that $[D_{X}, D_{Y}] = D_{\{Y,X\}}$.

The meaning of the expansion formula (\ref{BCH}) is that the propagator obtained by successive application
of $e^{\lambda A /2}$, $e^{\lambda B}$ and $e^{\lambda A/2}$ is a propagator of an equivalent Hamiltonian
$\tilde H$ equal to the original $H = A + B$, only in the limit of $\lambda \rightarrow 0$.
For a small enough time step $\lambda$, the exact solution under $\tilde H$ is expected to converge to a solution of $H$.
This propagator is symplectic and also preserves the symmetry, time reversibility and
first order invariants of the system.
When the perturbation $V$ does not depend on time then both Hamiltonians $H$ and $\tilde H$
are conserved. Therefore the global deviation from energy conservation is directly
accessible from $\tilde H$ - $H$.
This is a valuable asset of the symplectic methods, since most good-for-all integration
methods cannot easily predict the global behavior of their solutions, in general.

Explicitly, from Eq. (\ref{BCH}), on choosing $A = D_{V}$ and $B = D_{H_{0}}$,
one gets for the equivalent Hamiltonian
\[
{\tilde H} = H - \frac{\lambda^{2}}{24} \{\{H_{0},V\}, V + 2 H_{0}\} + {\cal O}(\lambda^{4}) .
\]
Hence this propagator is second order in the time step $\lambda$. The global error correct 
up the second order in $\lambda$ is then
\be
\label{error2}
\Gamma^{(2)} = {\tilde H} - H = -\frac{\lambda^{2}}{24}
\left( {\bf F}^{2} + 2 {\bf F}\cdot {\bf F}_{c} + 2 {\bf p}\cdot\frac{\partial}{\partial {\bf r}} 
({\bf p}\cdot {\bf F}) \right)
\ee
where $\bf F = - \partial V / \partial {\bf r}$ is the perturbation force and $F_{c} = -{\bf r}/r^{3}$
is the Coulomb force. For small values of $r$, the error is therefore dominated by 
$\lambda^{2} {\bf F}\cdot{\bf r} / r^{3}$. In contrast, the global error for
a $T + V$ splitting, obtained from Eq. (\ref{BCH}) and using $A = D_{V - 1/r}$ and $B = D_{T}$, is
\[
\Gamma^{(2)'} = \Gamma^{(2)} -\frac{\lambda^{2}}{24} \left( {\bf F_{c}}^{2} + 
2 {\bf p}\cdot\frac{\partial}{\partial {\bf r}} ({\bf p}\cdot {\bf F_{c}}) \right)
\]
The truncation error for the $T+V$ splitting is still second order in the time step,
but is clearly inferior at small distance $r$, where it behaves as $1/r^4$
 
Formula (\ref{BCH}) suggests a couple of ways to improve the performance of this method.
First, if operator $B$ is replaced by $B' = B + \lambda^{2}/24\;[A+2 B,[A,B]]$ and providing
the the evolution under $B'$ is explicitly known, then the $\lambda^{3}$ term in expansion 
is removed and the error is now of the order of $\lambda^{5}$.
The second way of accelerating the convergence is obtained
by adapting the times step. Superior order methods are obtained by composing steps with appropriate
variable time steps. Examples of such schemes are presented in the next section. 

A solution of the pertubed Kepler problem is therefore obtained by evolving the phase
space point $({\bf r}, {\bf p})$ of the system successively, under two elementary operations.
The ``drift stage'' is the evolution under the Kepler Hamiltonian $H_{0}$, while the evolution
under the perturbation Hamiltonian $V$ is called the ``kick stage''. 

\subsection{Drift stage}
Although a textbook example, an explicit solution of the Kepler problem is not entirely trivial.
To state the problem, given the position ${\bf r}_{0}$, and momentum ${\bf p}_{0}$ at time $t_{0}$, one
need to find positon $\bf r$ and momentum $\bf p$ at some other time $t_{0} + \Delta t$. This mapping
denoted by $D(\Delta t)$, gives the position and momentum along a Keplerian orbit, from an initial
position and momentum, after a time $\Delta t$.

The trajectory during the drift stage is a segment of an ellipse, a parabola or a hyperbola, 
depending whether the ``local'' energy $w = H_{0} = {\bf p}^{2}/2 - 1/r$ is 
positive, zero or negative, respectively.
The geometric size of the ellipse and the orbital period also depend on the local energy.
The angular momentum
${\bf L} = {\bf r}\times{\bf p}$ and the Runge-Lenz
${\bf A} = ({\bf p}^{2}-1/r)\, {\bf r} - ({\bf r}{\bf p})\,{\bf p}$
vectors, are used to identify the orientation of the orbital plane in space. 
Although $w$, ${\bf L}$ and ${\bf A}$ can be calculated from the initial state vector (${\bf r}_0,{\bf p}_0$),
it is advantageous to keep these quantities in the state vector, alongside with position and momentum.
This helps save a number of floating point operations.
Besides it can also help avoid the accumulation of round-off errors that might creep up in the calculation.
For instance the energy is sometimes obtained as a difference of two large numbers: the
kinetic and potential energies.

Having as input the thirteen dimensional state vector $({\bf r}, {\bf p}, w, {\bf L}, {\bf A})$,
the drift stage proceed as follows:
\begin{enumerate}
\item obtain the characteristic parameters of the orbit: semimajor axis $a = 1/2w$, eccentricity
$\epsilon = \|{\bf A}\|$ and orbital angular frequency $\omega = (2|w|)^{3/2}$,
\item calculate the direction of the pericenter $\hat e_{1} = {\bf A}/\epsilon$ and
the direction in the orbital plane perpendicular to it, $\hat e_{2} = {\bf L} \times {\bf A} /\epsilon L$,
\item find the eccentric anomaly corresponding to the initial position
$u_{0} = \arctan (1-2|w|r, \sqrt{2|w|}{\bf r}{\bf p})$,
\item find the eccentric anomaly $u$ after time $\Delta t$
as $u = \mbox{Kepler}(\epsilon, u_{0} - \epsilon\sin u_{0} + \omega \Delta t)$,
where $\mbox{Kepler}(\epsilon, M)$ is the solution of Kepler's equation $u - \epsilon \sin u - M = 0$
as a function of parameters $\epsilon$ and $M$,
\item calculate the new position on the orbit corresponding to the new eccentric anomaly $u$ as
\begin{eqnarray*}
{\bf r} &=&  a (\cos u - \epsilon)\,\hat e_{1} + a \sqrt{1-\epsilon^{2}} \sin u\,\hat e_{2}\\
{\bf p} &=& -\frac1{\sqrt{a}} \frac{\sin u}{1 - \epsilon \cos u}\,\hat e_{1} + \sqrt{\frac{1-\epsilon^{2}}{a}}
\frac{\cos u}{1-\epsilon \cos u}\,\hat e_{2}\;,
\end{eqnarray*}
\item energy, angular momentum and Runge-Lenz vectors are not modified during the drift stage.
\end{enumerate}

The steps above apply specifically to the case of negative energy (elliptic orbit). It is not difficult to
generalize this procedure for the parabolic and hyperbolic motions.

Up to round-off errors, the drift stage integrates exactly the orbit for any time step $\Delta t$, except for the
solution of the transcendental Kepler's equation which has to be obtained approximately. However, this equation
has been long and carefully studied, as Goldstein remarks \cite{Goldstein}:
\begin{quotation}
{\em Indeed, it can be claimed that the practical need to solve Kepler's equation to accuracies of a second of arc over the
whole range of eccentricity fathered many of the developments in numerical mathematics in the eighteenth and
nineteenth centuries. A few of the more than 100 methods of solution developed in the pre-computer era are
considered in the exercises to this chapter.}
\end{quotation}

\subsection{Kick stage}

In the kick stage the system evolves solely under the perturbation Hamiltonian $V$.
This mapping is denoted by $K(\Delta t)$.
If $V$ does not depend on momentum, then the position vector is a cyclic coordinate and does
not change during this stage. The change in momentum can then be explicitly obtained as
\[
{\bf p}' - {\bf p} = \Delta {\bf p} =  \int_{t_{0}}^{t_{0} + \Delta t} {\bf F} dt
\]
Where $\bf F$ is the perturbation force derived from the potential $V$.
Quantities $w$, ${\bf L}$ and ${\bf A}$ in the state vector are
updated from ${\bf r} $, ${\bf p}$ and $\Delta {\bf p}$, instead of calculating them
directly from ${\bf r} $ and ${\bf p}'$. This precludes the accumulation of round-off
errors and increases the efficiency of the procedure.
For example, energy is updated during the kick stage as
\[
w' = w + {\bf p}\Delta{\bf p} + \frac 12 \Delta {\bf p}^{2}
\]

\subsection{Kepler solver}

The solution of the transcendental Kepler's equation is the most time consuming part in the propagator.
The traditional numerical scheme to obtain accurate solutions is to ``guess'' a good starting approximation
and then refine it by using Newton-Raphson iterations until the desired accuracy is obtained. Each iteration
involves evaluating trigonometric functions several times. Since each trigonometric function evaluation has
a cost of at least several hundreds of microprocessor clocks, regardless if it is done ``on-the-chip'' or by
a library call, the cost of solving the Kepler's equation can mount easily to thousands of clocks. It is clear
that a long time integration, on the order of $10^{8}$ time steps would require a more refined procedure. A
table-driven procedure is proposed here, which trades memory space in favor of time. No trigonometric functions
are calculated during iterations.

The Kepler solver takes two branches, depending whether the orbit is elliptic or hyperbolic. For negative energy,
or when $0 \le \epsilon \le 1$, the equation to solve is
\be\label{kepler1}
u - \epsilon \sin u - M = 0 .
\ee
In the hyperbolic case, for positive energy, or for $1 \le \epsilon \le \infty$, Kepler's equation has
the form
\be\label{kepler2}
\epsilon \sinh u - u - M = 0.
\ee

Equation (\ref{kepler1}) is solved as follows:
\begin{enumerate}
\item using the fact that Eq. (\ref{kepler1}) is invariant to 
($M \rightarrow M + n \pi$, $u \rightarrow u + n \pi$, $\epsilon \rightarrow (-1)^{n} \epsilon$) 
and ($M \rightarrow \pi - M$, $u \rightarrow \pi - u$) transformations, the argument $M$ of the
equation can be mapped to the $[0,\pi)$ interval. The equation needs to be solved only for
$0 \le \epsilon \le 1$ and $0 \le M \le \pi$. The solution for arbitrary $M$ is obtained by
adding $n \pi$.

\item for large eccentricity ($\epsilon \rightarrow 1$) and low $M$ the solution of Eq. (\ref{kepler1})
has an essential singularity. This can be seen by seeking a solution as power series in $M$. Equating
the coefficients, order by order, one gets $ u(\epsilon, M) = M/(1-\epsilon) -M^{4}\epsilon/6 (1-\epsilon)^{4}+
{\cal O}(M^{5})$. This series does not converge uniformly when $\epsilon \rightarrow 1$. On the other hand
when $\epsilon = 1$ and the Taylor expansion of $\sin$ is used one gets $u(\epsilon = 1, M) = (6 M)^{1/3} + 
{\cal O}(M)$. A direct Newton-Raphson approach is started for the extreme cases when 
$\epsilon -1 \approx M \approx 0$, using $(6 M)^{1/3}$ as initial guess.

\item for the ``regular'' cases, a grid is set up for the $(0, \pi)$ interval with points $u_{k} = k \pi/n_{g}$,
$k = 0, \ldots, n_{g}$ and a table of sine and cosine at the grid points is stored at the start of the program.
From an initial grid index guess $u = M$, a Newton-Raphson iteration for indices is started using the following mapping:
$k \rightarrow k' = k + \Xi\big[(M + \epsilon s_{k} - u_{k})/(1 - \epsilon c_{k})\big]$, where $\Xi$ is a
function returning the integer truncation of a real number, and $s_{k}$ and $c_{k}$ are the sine and cosine values at the
grid point $k$. At the end of this process the solution is localized between
two grid points with an accuracy no greater than the grid spacing $\pi/n_g$. In this way, the process has
both the quadratic convergence of Newton-Raphson's method and the stability of bisection method.

\item A last fifth-order Newton-Raphson refinement is obtained using the grid point closest to the solution
obtained at the previous step. However, instead of searching for
a $u$ value which satisfies Eq. (\ref{kepler1}), it is more efficient to look for a solution in the unknown $\tan(u/2)$.
This procedure delivers then directly the pair $(\sin u, \cos u)$ corresponding to the solution, and hence no
trigonometric function need to be explicitly calculated during the procedure call! Assuming that the
values of the function $f_{0}$ and of the first four derivatives $f_{1}$, $f_{2}$, $f_{3}$ and $f_{4}$ are known
at the grid point, then the following corrections are obtained
\begin{eqnarray*}
\delta_{2} &=& \frac{f_{0}}{f_{1}} \quad \delta_{3} = \frac{f_{0}}{f_{1} + \frac{\delta_{2}}2 f_{2}}\\
\delta_{4} &=& \frac{f_{0}}{f_{1} + \frac{\delta_{3}}2 f_{2} + \frac{\delta_{3}^{2}}6 f_{3}} \\
\delta_{5} &=& \frac{f_{0}}{f_{1} + \frac{\delta_{4}}2 f_{2} + \frac{\delta_{4}^{2}}6 f_{3} + \frac{\delta_{4}^{3}}{24}f_{4}}
\end{eqnarray*} 
to give the approximate solution $\tan u/2 \approx \sqrt{(1-c_{k})/(1+c_{k})} + \delta_{5}$

\end{enumerate}

The accuracy of this procedure is expected to be of the order $(\pi/n_{g})^{5}$. Indeed, in tests using $n_{g} = 1024$
the error in finding both $\sin u$ and $\cos u$ where $10^{-14} -- 10^{-15}$ except for the range of
$\epsilon \rightarrow 1$ and $M \rightarrow 0$ where the error is no greater than $10^{-12}$.

A similar approach is taken to solve Eq. (\ref{kepler2}). Although Kepler's equation for positive energy is not
manifestly periodic, a mapping to the standard interval is obtained by observing that Eq. (\ref{kepler2}) can
be written as $\epsilon_{1} e^{u} - \epsilon_{2} e^{-u} - u - M = 0$ and that the following set of transformations
\[
u \rightarrow u + n D \quad M \rightarrow M + n D \quad \epsilon_{1,2} \rightarrow \epsilon_{1,2} e^{\pm n D}
\]
leave this form invariant. In this way, the equation needs to be solved only in the standard interval $[0, D)$ where
$D$ is arbitrary. For convenience, $D$ is chosen $D = 2.0$. This interval is gridded and the exponential 
$e_{k} = \exp u_{k}$ is  calculated at each grid point $k$. The Newton-Raphson grid iteration
\[
k \rightarrow k' = k + \Xi\big[(M + u_{k} - \epsilon_{1} e_{k} - \epsilon_{2} e_{k}^{-1})/
(\epsilon_{1} e_{k} + \epsilon_{2} e_{k}^{-1} - 1)\big]
\]
ends by identifying the grid point closest to the solution. The solution for $e^{u}$ is
obtained from this grid point after a fifth order Newton-Raphson refinement. This subroutine returns
an approximation for the $(\sinh u, \cosh u)$. Precision levels similar to the elliptic case are
obtained.

When tables of sine, cosine and exponential are build at the set up of the program, the drift stage
can then be performed without evaluation of any transcendental function, which brings significant improvements
to the efficiency of the integrator.

\section{Applications}
\subsection{Kepler atom in uniform electric field}

In the absence of a perturbation, the kick stage reduces to identity and the dynamics is
described only by the drift. The propagator was tested and the solution was seen to be
practically exact (within the machine precision) even for extremely long time integration.
A proper test of this symplectic integrator can only be done in the presence of a perturbation.
The simplest perturbation, which is also completely integrable, is the constant and uniform
force field.

When a constant force ${\bf F}$ acts on the system, the Kepler orbit orbit starts to precess and change
its eccentricity. The corresponding potential is $V = - {\bf r}{\bf F}$, such that the
energy $E = {\bf p}^{2}/2 - 1/r - {\bf r}{\bf F}$ is conserved. For forces $F > (-E/2)^{2}$
the system can break away and ionize.

The angular momentum and the Runge-Lenz vectors evolve in
time according to
\[
\frac{d {\bf L}}{dt} = {\bf r}\times {\bf F}
\quad\quad
\frac{d}{dt} \left[ {\bf A} - \frac 12 {\bf r}\times ({\bf r} \times {\bf F})\right] = 
\frac 32 {\bf F}\times{\bf L}
\]
If the orbit does not change appreciably over one period, then the average angular momentum and
Runge-Lenz vectors obey equations
\[
\frac{d}{dt}\langle L\rangle = \frac 32 {\bf F}\times \langle {\bf A}\rangle \quad\quad
\frac{d}{dt}\langle A\rangle = \frac 32 {\bf F}\times \langle {\bf L}\rangle
\]
because of the Pauli's replacement rule ${\bf r} \rightarrow - (3/2) {\bf A}$, and the fact 
that ${\bf r} \times ({\bf r}\times{\bf F})|_{0}^{T} \approx 0$.
Within these assumptions, the slow changes in $\bf L$ and $\bf A$ are obtained, by solving the above
system of equations, as
\be\label{sol1}
\langle {\bf L}\rangle = \cos(\frac 32 F t) {\bf L}(0) + [1 - \cos(\frac 32 F t)][\hat F\cdot{\bf L}(0)]\hat F +
\sin(\frac 32 F t) [\hat F \times {\bf A}(0)]
\ee
\be\label{sol2}
\langle {\bf A}\rangle = \cos(\frac 32 F t) {\bf A}(0) + [1 - \cos(\frac 32 F t)][\hat F\cdot{\bf A}(0)]\hat F +
\sin(\frac 32 F t) [\hat F \times {\bf L}(0)] 
\ee
Both $\bf L$ and $\bf A$ therefore rotate about each other with a period of $4 \pi/3 F$.

The simplest second order symplectic integrator ({\em step2})
\[
S(\Delta t) = K(\Delta t/2) D(\Delta t) K(\Delta t/2)
\]
requires only one ``drift'' stage $D$ and two ``kick'' operations $K$. The ``drift'' stage involves
solving Kepler's equation and has a higher computational cost than the ``kick'' stage. An equivalent
integrator $DKD$ is not as efficient, because it uses two ``drift'' stages. 

\begin{figure}
\begin{center}
\includegraphics[width = 4.5in]{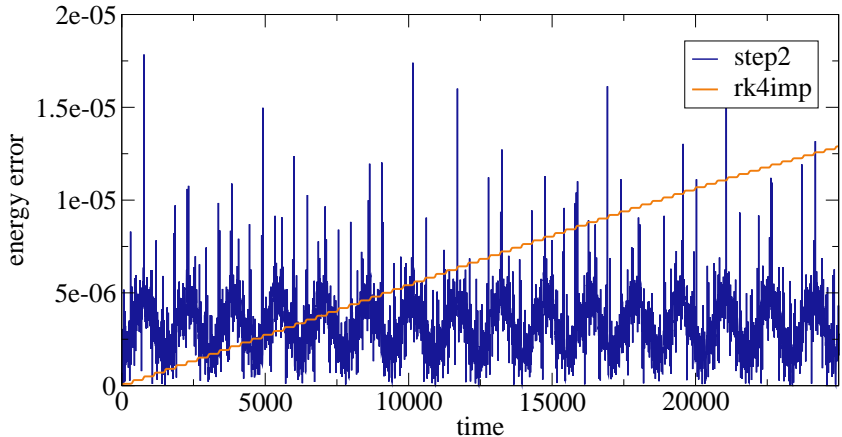}
\end{center}
\caption{\label{longtime}
The relative energy error is compared for long time integration using the step2 integrator and implicit
Runge-Kutta of order 4 from GSL. The initial orbit has eccentricity $\epsilon = 0.9$ and energy $w = -0.5$. The
electric field is $5.5\times 10^{-3}$ oriented perpendicularly on the orbit plane. The fixed time step for
{\em step2} corresponds to 200 steps per orbit.
}
\end{figure}

Figure \ref{longtime} compares the performance of {\em step2} integrator with
a standard implicit Runge-Kutta method of order 4,
with adaptive time step ({\em rk4imp}), from the free GNU Scientific Library (GSL) \cite{GSL}. The initial orbit
has eccentricity 0.9, energy -0.5 and period $2\pi$ in the chosen units.
An uniform electric field of magnitude $5.5\times10^{-3}$ is applied along
a direction perpendicular to the orbital plane. If the initial orbit represents a ground state
hydrogen atom, then the electric field is $2.86\times 10^{7}$ V/cm. Because of the scaling of
the classical equations of motion, this would also simulate a $n=100$ Rydberg atom in a field
of intensity 2.86 kV/cm. The trajectory is simulated for about 4000 orbits, until time 25000.
The {\em step2} subroutine takes $8\times 10^{5}$ steps of constant size $\pi/100$ and finishes
the jobs in 0.7 seconds, while {\em irk4imp} makes $7.6\times 10^{6}$ steps and takes 9.6 seconds to
complete. The precision and accuracy parameters are set to $10^{-5}$. Although the standard Runge-Kutta
integrator runs ten times longer and makes ten times more steps, the relative error for the energy
conservation increases. The performance of {\em step2} are initially worse than {\em irk4im}, but the
accumulation of errors is much slower. The long time integration advantages of the symplectic method
are clear.

\begin{figure}
\begin{center}
\includegraphics[width = 2.5in]{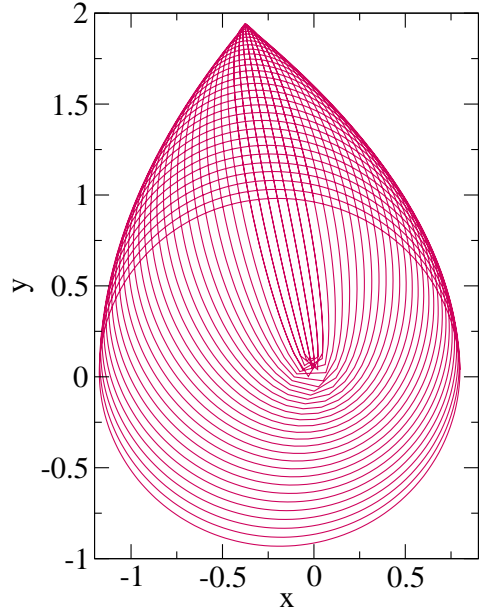}
\end{center}
\caption{\label{orbit}
The trajectory when electric field of magnitude $5.5\times 10^{-3}$ is oriented parallel to the orbit plane.
When the eccentricity becomes 1 the traditional integrators fail because the particle comes arbitrarily close
to the force center. The symplectic integrator is able to ``gracefully'' go through this singularity.
}
\end{figure}

As shown by Eqs. (\ref{sol1}) and (\ref{sol2}), when the electric field is oriented parallel to the orbital plane,
the trajectory goes to a singular orbit with $L = 0$ and unit eccentricity,
and the particle goes through the Coulomb center.
The error accumulates at much higher rate for the standard Runge-Kutta integrator; every time the
singularity is encountered, the error increases at least one order of magnitude. A fragment of a
trajectory having this kind of singularity is shown in figure \ref{orbit}.
Owing to the built-in exact Kepler solution, the symplectic integrator is able to advance through 
this singularity with no catastrophic consequences. 
At very small distances, the central Coulomb force is much more
stronger than the external field and the dynamics is practically governed by the drift stage alone.
In order to cope with such extreme situations, the basic {\em step2} integrator can be 
improved in several ways.

Higher order integrators can be obtained by compounding more stages during one time step. Following \cite{Yoshida1990},
a fourth order {\em step4} integrator is obtained using the following symmetric sequence:
$K_{1} D_{1} K_{2} D_{2} K_{2} D_{1} K_{1}$, which uses only three drift stages. Here the following notation is used:
$K_{1,2} = K(a_{1,2}\Delta t)$ and $D_{1,2} = D(a_{1,2}\Delta t)$.
A sixth order ({\em step6}) stepping procedure is obtained by using an appropriate combination of 
elementary {\em step2} steps. For example \cite{Yoshida1990}, the following sequence 
$S_{3} S_{2} S_{1} S_{0} S_{1} S_{2} S_{3}$
has an error of order 6 in the time step, is symmetric and time reversible. Here $S_{0,1,2,3}$ means {\em step2}
steps with time steps $w_{0,1,2,3}\times\Delta t$. The coefficients $w_{0,1,2,3}$ are solutions of a nonlinear order
equation which ensure that all errors up to order 6 are canceled.
The numerical coefficients used in these higher order integrators are listed in Table \ref{coeffs}.

\begin{table}
\caption{\label{coeffs}
Numerical coefficients used in optimized higher order symplectic
schemes.}
\begin{center}
\begin{tabular}{ll}
\hline
$a_{1} =  0.6756035959798288$ & $a_{2} = -0.17560359597982883$ \\
$b_{1} =  1.3512071919596578$ & $b_{2} = -1.7024143839193149$ \\
\hline
$w_{0} =  1.3151863206839063$ & $w_{1} =   -1.17767998417887$ \\
$w_{2} =   0.235573213359357$ & $w_{3} =   0.784513610477560$ \\
\hline
\end{tabular}
\end{center}
\end{table}

Figure \ref{checkS} compares the performance of {\em step2}, {\em step4} and {\em step6} routines as a function of
the time step for eight orbits, for low and high eccentricity orbits. Electric field of strength $5.5\times 10^{-3}$
is oriented parallel to the orbital plane.
As expected, higher order integrators have the error decreasing faster with decreasing time steps.
However, this behavior is evident only when the
time step is smaller than a critical time step, which decreases with increasing eccentricity. For eccentricity
0.9, the advantage of higher orders is manifest only for time steps smaller than $10^{-2}$, for example.
In order to understand this feature it is enough to consider the leading terms of expansion (\ref{BCH}) in
evaluating the global error:
\be\label{dH}
{\tilde H} - H \approx -\frac{\lambda^{2}}{12}\frac{F}{r^{2}}  + \frac{\lambda^{4}}{720}\frac{F}{r^{5}} + \ldots
\ee
The maximum error is obtained when $r$ has a minimum at $r \sim 1 - \epsilon$.
The expected convergence is obtained when the $\lambda^{4}$ correction is smaller than the $\lambda^{2}$ one,
or when $\lambda  < \sqrt { 6 (1-\epsilon)^{3}}$. Indeed, for $\epsilon = 0.4$ one gets $\lambda < 1.1$, and
for $\epsilon = 0.9$ one gets $\lambda < 0.07$, roughly in agreement with the results presented in 
Fig. {\ref{checkS}. The error saturates around $10^{-12}$ because of the limited precision 
imposed by the Kepler solver. A finer grid in the Kepler solver improves this precision.

\begin{figure}
\begin{center}
\includegraphics[width = 2.5in]{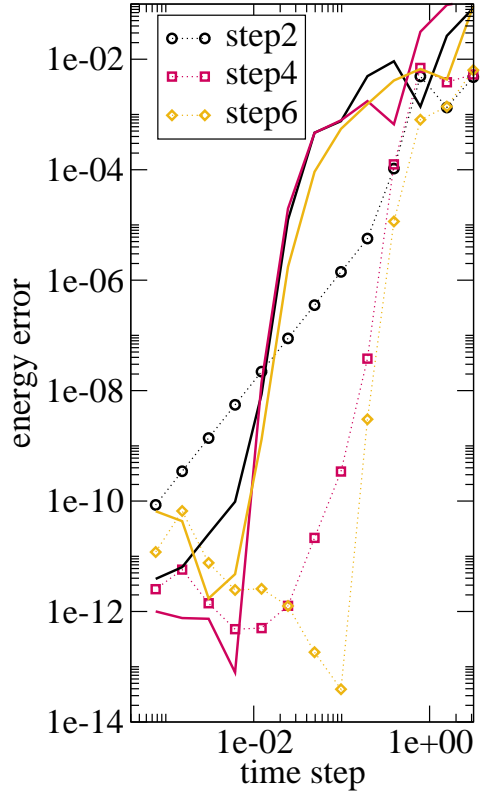}
\end{center}
\caption{\label{checkS}
Relative error in energy after eight orbits when {\em step2}, {\em step4} and {\em step6} are used.
Both low eccentricity ($\epsilon = 0.4$) starting orbit, with symbols and dotted lines, and highly 
eccentric orbit ($\epsilon = 0.9$), with solid lines, are represented.}
\end{figure}

The error in energy for the basic integrator {\em step2} becomes unbounded in the case of a field
parallel to the orbital plane, as the eccentricity becomes unity with a period given by $4\pi/3 F$.
Another way of improving the basic integrator {\em step2} is to use an adaptive time step strategy.
The singularity is removed from the first term in Eq. (\ref{dH}), and weakened for the second one,
if the time step $\lambda$ is adaptively chosen proportional to distance $r$ as $\lambda = \eta r$.
Figure \ref{TA} shows the results using this strategy ({\em stepA}). 
The electric field has a strength of $\pi/600$ so that that orbit become singular with
a period of 800, as predicted. Eccentricity, as plotted in the lower graph in Fig. \ref{TA}, is
initially 0.2. The energy conservation is bounded, in general, and has spikes whenever eccentricity
goes to 1 and the orbit becomes one dimensional, because of the $\lambda^{4}$ (and higher) 
energy correction which dominates in these cases.
In contrast, the implicit Runge-Kutta ({\em rk4imp})
shows a catastrophic accumulation of error after the first encounter with singularity, even though 
the precision and the accuracy parameters are set at $10^{-8}$ and about $10^{6}$ steps are taken for
the segment shown in figure. About the same number of steps are taken by {\em stepA} to integrate
the orbit over the whole time interval.
\begin{figure}
\begin{center}
\includegraphics[width=4.5in]{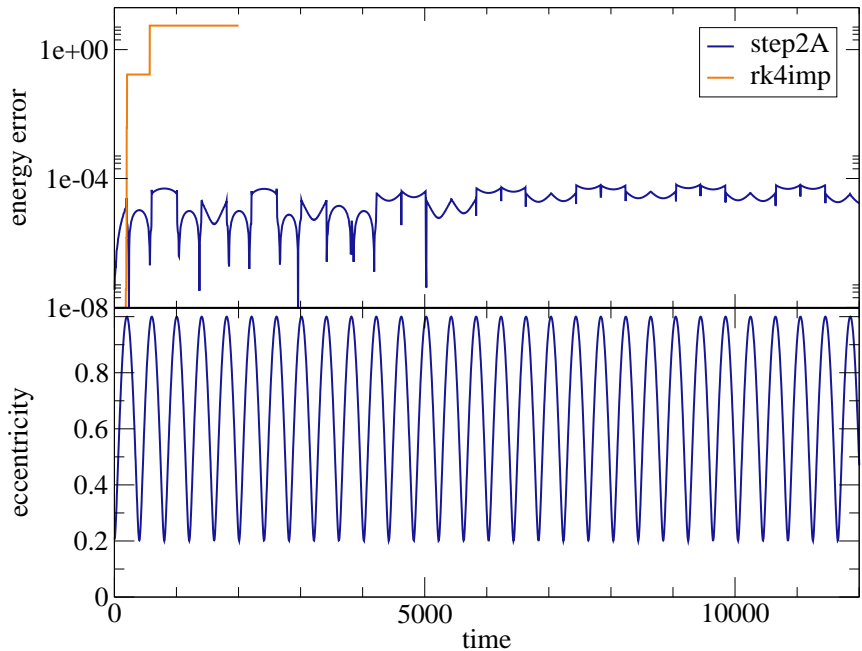} 
\end{center}
\caption{\label{TA}
The relative error in energy for an orbit integrated with a time adaptive step (upper graph).
Eccentricity as shown in the lower graph, goes periodically to unity because the electric field is parallel to
the orbital plane.}
\end{figure}

\subsection{Kepler atom in monochromatic time dependent field}

Time dependent fields can be taken into account by using a canonical transformation which adds time
as a position coordinate to the Hamiltonian (see for example \cite{Candy1991}). Therefore, only during
the drift stage the time variable is advanced.
The time dependent external force does work on the system and the energy is not conserved.
However, when the work is subtracted, the
quantity
\[
w - \int_{0}^{t} {\bf F}(s) {\bf v}\; dt 
\]
is conserved, and can be used to quantify the precision level of the integrator.

Figure \ref{mono} show the results from a long time integration of Kepler orbit that was started with energy -0.5 and
eccentricity 0.9, under a monochromatic uniform field of magnitude 0.1, frequency 2.2 and orientation perpendicular
to the orbital plane.
During one time step,
the trajectory is evolved according to the scheme ({\em step2T} time dependent version of {\em step2}):
K($\Delta t/2$) F(t) D($\Delta t$) K($\Delta t/2$), where D and K represent the drift and kick stages, while
F is the force calculation step at time $t_{0} + \Delta t$. Here $t_{0}$ is the time at the beginning of
the step. The orbit is advanced for $3\times10^{6}$ time steps of size $\pi/100$. Figure \ref{mono} shows
the variation of energy (upper graph) and the deviation from conservation of the energy-minus-work quantity
(lower graph) for the last segment of the run. 

\begin{figure}
   \centering
   \includegraphics[width=4.5in]{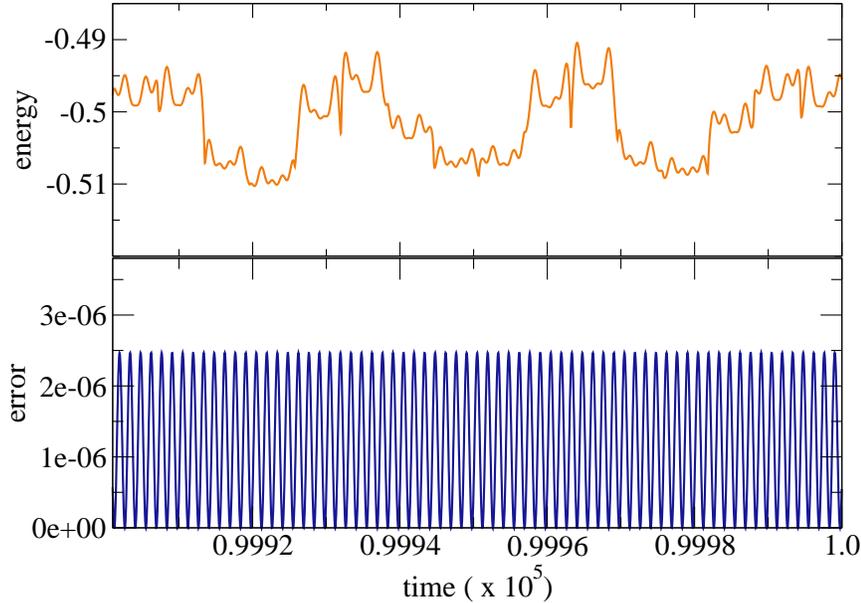} 
   \caption{Atom energy (upper graph) and the relative error of the energy minus the work done by the field
   (lower graph) for the last segment of a long time run.}
   \label{mono}
\end{figure}

\section{Conclusion}

The integration of the orbit of a particle having a perturbed Keplerian motion can take advantage of the
explicit integrability of the Kepler problem. Splitting the Hamiltonian as a Kepler part plus a 
perturbation, as opposed to the more general kinetic plus potential energy splitting, has clear
advantages, especially for long time integration.

The overall efficiency of this method is limited by how fast the transcendental Kepler equation
can be solved. By using a table of pre-calculated trigonometric and exponential functions, and
a fifth order Newton-Raphson refinement, a fast and reasonably accurate Kepler solver is
successfully used.

The convergence of the second order, basic integrator {\em step2}, can be improved to obtain fourth order
{\em step4} and sixth order {\em step6} symplectic schemes. A time adaptive step {\em stepA} has been 
proved to have excellent energy conservation for long time integration, when the trajectory goes
repeatedly through the Coulomb singularity, in the case of an uniform constant field in the orbital plane.
Time dependent problems can also be solved using a variant ({\em stepT}) of the basic step, as demonstrated
for a monochromatic field.

\section*{Acknowledgments}
This work was carried out under the auspices of the U.S. Department of Energy at Los Alamos National Laboratory 
under Contract No. DE-AC52-06NA25396.

\end{document}